\documentclass[runningheads]{llncs}
\usepackage{color}
\usepackage{xcolor}
\usepackage{graphicx}
\usepackage{multirow}

\usepackage{caption}
\usepackage{datatool}
\usepackage{subcaption}
\usepackage{xcolor}
\usepackage{tikz}
\usepackage{enumerate}
\usepackage{algorithm}
\usepackage{algpseudocode}
\usepackage{adjustbox}
\usepackage{hyperref}
\usepackage{verbatim}
\usepackage{amsmath}

\hypersetup{
	colorlinks=true,
	citecolor=blue 
}

\newcommand{\methodname}{FeatPCA}

\begin{document}

\title{{\methodname}: A feature subspace based principal component analysis technique for enhancing clustering of single-cell RNA-seq data}

\author{
    Md Romizul Islam, Swakkhar Shatabda 
}

%
%
%
%
\institute{}
\maketitle              
	\begin{abstract}
		Single-cell RNA sequencing (scRNA-seq) has revolutionized our ability to analyze gene expression at the cellular level. By providing data on gene expression for each individual cell, scRNA-seq generates large datasets with thousands of genes. However, handling such high-dimensional data poses computational challenges due to increased complexity. Dimensionality reduction becomes crucial for scRNA-seq analysis. Various dimensionality reduction algorithms, including Principal Component Analysis (PCA), Uniform Manifold Approximation and Projection (UMAP), and t-Distributed Stochastic Neighbor Embedding (t-SNE), are commonly used to address this challenge. These methods transform the original high-dimensional data into a lower-dimensional representation while preserving relevant information. In this paper we propose {\methodname}. Instead of applying dimensionality reduction directly to the entire dataset, we divide it into multiple subspaces. Within each subspace, we apply dimension reduction techniques, and then merge the reduced data. {\methodname} offers four variations for subspacing. Our experimental results demonstrate that clustering based on subspacing yields better accuracy than working with the full dataset. Across a variety of scRNA-seq datasets, {\methodname} consistently outperforms existing state-of-the-art clustering tools.
	\end{abstract}

		\section{Introduction}
	Single-cell RNA sequencing technology is aiding us to have a better understanding of gene expression at single cell level. The technology allows us to measure the mRNA expression of every gene for each cell.  This capability is incredibly powerful because it helps us understand which genes are turned on, how much they’re expressed, and where they’re active within the cell \cite{seq1}. This technology has opened many windows for us to analyze the cells and genes expression \cite{seq2}. Single-cell data plays a pivotal role in cancer research, multi-omics studies, genome assembly, and various other aspects of biology and bioinformatics\cite{applications1}\cite{applications2}\cite{applications3}. Here, we have a cell-by-gene matrix and each cell of the matrix corresponds to the gene expression or mRNA count for a cell for a specific gene, thus giving us cellular-level data.\\\\In single-cell data, missing values (representing absent gene expression) are common\cite{missingvalues}. To address this, data imputation becomes necessary to recover these missing values\cite{imputing}. One effective approach is using autoencoders, which learn a compact representation of the data and can predict missing entries based on existing information\cite{autoimpute}. The single-cell datas are very high-dimensional. High-dimensional scRNA-seq data are typically normalized and projected to a lower-dimensional space by dimension reduction\cite{dimred1}\cite{dimred2}. A core analysis of the scRNA-seq transcriptome profiles is to cluster the single cells to reveal cell subtypes and infer cell lineages based on the relations among the cells\cite{clustering1}\cite{clustering2}. Clustering is a necessary step to identify the cell subpopulation structure in scRNA-seq data. Different algorithms such as K-means, Hierarchical clustering are generally used to cluster the cells\cite{kmeans}. Clustering algorithms are typically applied on the lower-dimensioned data to find the clusters. Dimensionality significantly impacts the performance of clustering algorithms in bioinformatics. As the number of features (dimensions) increases, the data space becomes sparser. Sparse data is challenging for clustering because reliable density estimates are harder to obtain. High-dimensional data can lead to instability and decreased classification accuracy in clustering\cite{dimensionality}. Clustering high-dimensional data is computationally expensive\cite{dimensionality1}. Different dimension reduction techniques such as PCA, UMAP\cite{umap}, t-SNE\cite{tsne}, LDA\cite{lda}, autoencoders\cite{autoencoders} etc. are typically used to reduce the dimensions of the data. Sometimes, the data points are very large and distributed over different servers. PCA is applied locally on each of the data divisions separately and later they get combined together\cite{d1}. These projected datas on the lower dimension can act as a proxy for the original data in k-means clustering\cite{d2}.
	\
	\\\\ 
	 The method we propose, called {\methodname}, partitions the feature sets into multiple slices rather than dividing the data points themselves. The process of division is described in detail in Section \ref{sec:division}. By doing so, we can more effectively analyze and process the features, gaining a deeper understanding of the underlying patterns. {\methodname} partitions the feature set, resulting in multiple smaller subsets of features. In our case, the feature sets denote the genes. Generally principle component analysis (PCA) is done on the whole dataset to reduce the dimension. Here, our method applies PCA after dividing the dataset into several pieces. {\methodname} employs dimension reduction techniques on each of the smaller subsections. These principal components serve as a reduced representation of the original data, capturing the most significant variance within each subset. FeatPCA then combines all the principal components or embeddings to create a representation that represents the entire dataset. This merged dataset, now transformed and dimensionally reduced. By applying PCA to each gene subset and then integrating the results, we ensure that the most informative features from each division contribute to the final analysis. Then {\methodname} runs clustering algorithms on the merged data. {\methodname} also employs denoising autoencoders for data imputation and subsequently applies the same algorithm to the imputed data. This approach leads to superior performance compared to existing state-of-the-art clustering methods.

\section{Related Work}
 In the context of single-cell data, imputation is a common step to handle missing values. Various algorithms are employed for this purpose, and one effective approach is using autoencoders\cite{autoimpute1}. Autoencoders learn the underlying structure of the data and predict missing values based on the available information.  Dimensionality reduction methods play a crucial role in single-cell analysis. PCA is commonly employed to reduce dimensionality by calculating the principal components. These components capture the most significant variation in gene expression across cells, allowing for efficient representation and visualization of the data\cite{pcaapplication}. The Uniform Manifold Approximation and Projection (UMAP) method is commonly employed to reduce dimensionality and visualize single-cell data\cite{umapapplication}. t-Distributed Stochastic Neighbor Embedding (t-SNE) is a dimensionality reduction technique commonly used in single-cell RNA sequencing analysis while preserving local structure\cite{tsneapplication}. Autoencoder is also used to reduce dimensions\cite{autoencoderapplication}. Data partitioning is a fundamental technique in machine learning, particularly when dealing with large datasets\cite{datap1}. In certain scenarios involving large datasets distributed across multiple servers, the data is divided into segments. Each segment is processed independently on its respective server\cite{d1}. Within each segment, dimensionality reduction techniques are applied to reduce the number of features while preserving relevant information. After dimensionality reduction, the reduced data from all segments is combined, serving as a proxy for the original dataset\cite{d2}. This combined reduced data is then used for clustering purposes. When dealing with large datasets and distributed clustering of high-dimensional, heterogeneous data, a technique called Collective PCA can be employed. Collective PCA is specifically designed for distributed scenarios and can be used independently of clustering algorithms. It aims to reduce the dimensionality of the data while preserving essential information\cite{d3}.

\section{Proposed Method}
A simplified diagram of our proposed methodology is shown in Figure \ref{pic:pipeline}. As illustrated in Figure \ref{pic:pipeline}, our proposed method {\methodname} takes scRNA-seq data with a dimension of $n \times d$ as input, where $n$ is the number of cells and $d$ is the number of genes (features). This data is first normalized and subsequently, SCANPY \cite{scanpy} is used to find the highly variable genes, leading to dimension reduction of the genes from ($d$) to ($d'$).  {\methodname} then performs a critical step in the analysis by dividing the data based on its features. The features of the data is partitioned into k equal parts while keeping the cell dimension $n$ the same. The dimensions of the features in each partition become $m=\frac{d'}{k} + o$, where $o$ is the overlapping size. Then {\methodname} applies PCA on each subspace of the data to capture the most variance from each data partition. It was observed that dividing the data into $m$ partitions can help PCA emphasize specific feature sets that capture more meaningful representations, ultimately leading to enhanced clustering accuracy.  {\methodname} then merges the whole data into one section by concatenating the smaller partitions. Finally, {\methodname} applies existing clustering algorithms on the merged data. It was observed that the most variance can be captured for $k$ values from 2 to 20 across all standard scRNA-seq datasets. This is because partitioning the datasets into more segments may emphasize less useful information, which can further decrease clustering performance. 

\begin{minipage}{\textwidth}
		\begin{adjustbox}{width=\textwidth}
			\begin{tikzpicture}
				
				\fill[lightgray, rounded corners=8pt] (0cm,0cm) rectangle (2.5cm,4cm);
				\node[anchor=south west] at (0cm,2cm) {Input data};
				\node[anchor=south west, text width=2.5cm] at (0, 5) {\textbf{Normalize}};
				\node[anchor=south west, text width=2.5cm] at (1, 4) {d};
				\node[anchor=south west, text width=2.5cm] at (-.5, 2) {n};
				
				\draw[thick,->] (2.5,2) -- (3.5cm,2);
				
				\fill[olive,rounded corners=8pt] (3.5cm,0cm) rectangle (6cm,4cm);
				\node[anchor=south west, text width=2.5cm] at (3.5, 2) {Normalized data};
				\node[anchor=south west, text width=2.5cm] at (3.5, 5) {\textbf{Highly Variable gene selection}};
				\node[anchor=south west, text width=2.5cm] at (4.5, 4) {d};
				
				\draw[thick,->] (6, 2) -- (7cm,2);
				
				\fill[teal,rounded corners=8pt] (7cm,0cm) rectangle (9.5cm,4cm);
				\node[anchor=south west, text width=2.4cm] at (7, 1.5) {Data containing variable genes};
				\node[anchor=south west, text width=2.5cm] at (7, 5) {\textbf{Subspacing}};
				\node[anchor=south west, text width=2.5cm] at (8, 4) {d'};
				
				\draw[thick,->] (9.5, 2) -- (10.5cm,2);
				
				\fill[cyan,rounded corners=8pt] (10.5cm,0cm) rectangle (11.5cm,4cm);
				\node[anchor=south west, text width=2.5cm] at (10.6, .4) {\rotatebox{90}{Subspaced data}};
				\node[anchor=south west, text width=2.5cm] at (10.3, 4) {$\frac{d'}{k} + o$};
				
				\fill[cyan,rounded corners=8pt] (11.7cm,0cm) rectangle (12.7cm,4cm);
				\node[anchor=south west, text width=2.5cm] at (11.8, .4) {\rotatebox{90}{Subspaced data}};
				\node[anchor=south west, text width=2.5cm] at (11.7, 4) {$\frac{d'}{k} + o$};
				
				\draw[dashed,-] (12.7, 2) -- (13cm,2);
				
				\fill[cyan,rounded corners=8pt] (13cm,0cm) rectangle (14cm,4cm);
				\node[anchor=south west, text width=2.5cm] at (13.1, .4) {\rotatebox{90}{Subspaced data}};
				\node[anchor=south west, text width=2.5cm] at (13, 4) {$\frac{d'}{k} + o$};
				
				\node[anchor=south west, text width=2.5cm] at (11.5, 5) {\textbf{PCA}};
				
				\draw[thick,-] (14, 2) -- (14.5cm,2);
				\draw[thick,-] (14.5, 2) -- (14.5cm,-3);
				
				\fill[orange,rounded corners=8pt] (10.5cm,-5cm) rectangle (11.5cm,-1cm);
				\node[anchor=south west, text width=2.5cm] at (10.6, -4.6) {\rotatebox{90}{Top components}};
				\node[anchor=south west, text width=2.5cm] at (10.6, -5.5) {m1};
				
				\fill[orange,rounded corners=8pt] (11.7cm,-5cm) rectangle (12.7cm,-1cm);
				\node[anchor=south west, text width=2.5cm] at (11.8, -4.6) {\rotatebox{90}{Top components}};
				\node[anchor=south west, text width=2.5cm] at (11.86, -5.5) {m2};
				
				\draw[dashed,-] (12.7, -3) -- (13cm,-3);
				
				\fill[orange,rounded corners=8pt] (13cm,-5cm) rectangle (14cm,-1cm);
				\node[anchor=south west, text width=2.5cm] at (13.1, -4.6) {\rotatebox{90}{Top components}};
				\node[anchor=south west, text width=2.5cm] at (13.1, -5.5) {mk};
				
				\node[anchor=south west, text width=2.5cm] at (11.5, -1) {\textbf{Merge}};
				
				\draw[thick,->] (14.5, -3) -- (14cm,-3);
				
				\draw[thick,->] (10.5, -3) -- (9.5cm,-3);
				
				\fill[yellow,rounded corners=8pt] (7cm,-5cm) rectangle (9.5cm,-1cm);
				\node[anchor=south west, text width=2.5cm] at (7, -3) {{Merged data}};
				\node[anchor=south west, text width=2.5cm] at (7, -1) {\textbf{Clustering}};
				\node[anchor=south west, text width=3cm] at (7, -5.6) {m1+m2 ..+mk};
				
				\draw[thick,->] (7, -3) -- (6cm,-3);
				
				\draw[blue,rounded corners=8pt] (0cm,-5cm) rectangle (6cm,-1cm);
				
				\fill[blue] (1.2,-2) circle (.7cm) node[white] {Type 1};
				\fill[cyan] (3.2,-2.5) circle (.7cm) node[white] {Type 2};
				\fill[magenta] (5.2,-2.1) circle (.7cm) node[white] {Type 3};
				\fill[gray] (1.2,-3.8) circle (.7cm) node[white] {Type 4};
				\fill[violet] (5,-3.6) circle (.7cm) node[white] {Type 5};
			\end{tikzpicture}
			
		\end{adjustbox}
		\captionof{figure}{Pipeline of the {\methodname} algorithm}
		\label{pic:pipeline}
	\end{minipage}

		\subsection{Data Preprocessing}
		
	We perform standard preprocessing on the scRNA-seq dataset that includes: log-normalization and highly variable gene selection. SCANPY \cite{scanpy} is used to apply normalization and determine the highly variable genes from the dataset. Only the top highly variable genes from the normalized input data are kept in the dataset. The new dimension of the dataset becomes $n\times d'$, where $d'$ is the number of highly variable genes. We chose top 10000 genes from the highly variable genes across all our datasets.

                \subsubsection{Data Imputation} 
                The data containing highly variable genes has missing values in it. {\methodname} uses denoising autoencoder to replace the missing values. TENSORFLOW is used to build the autoencoder. The bottleneck layer contains 50 neurons in our analysis. As illustrated in the figure, {\methodname} uses the values got from the reconstructed input to impute the missing values in the original data.

		\subsection{Subspace Generation}
		\label{sec:division}
		The most crucial part of our proposed method is generating subspaces based on the features. {\methodname} divides the features into multiple segments and according to that whole dataset is divided into k partitions, where k varies from 2 to 20. For $k = 2$, the dataset gets divided into 2 parts. Subspace generation consists of 4 different approaches for determining meaningful representations from the processed datasets which are given as follows:

		\begin{enumerate}[i.]
			\item \textbf{Subspacing of the genes into equal parts sequentially :} 
                \label{method:normal}
			
			\begin{algorithm}
				\caption{Sequential Overlapping Subspace}
				\label{alg:normal}
				\begin{algorithmic}[1]
					\Procedure{SequentialOverlappingDivision}{FeatureSet}
					\State Define NumberOfPartitions, PartitionSize, OverlapSize
					\State Initialize StartIndex to 0
					\While{StartIndex $<$ Length(FeatureSet)}
					\State EndIndex $\gets$ StartIndex + PartitionSize
					\State DividedFeatureSets.add(FeatureSet[StartIndex:EndIndex])
					\State StartIndex $\gets$ StartIndex + (PartitionSize - OverlapSize)
					\EndWhile
					\EndProcedure
				\end{algorithmic}
			\end{algorithm}
			
		Approach-1 in	{\methodname} divides the data into k equal parts sequentially based on the features. This approach creates k subsets of data from the original data. During partitioning, overlapping is performed between adjacent partitions. The degree of overlapping depends on the partition size, with the algorithm permitting 20\% to 30\% of the partition size as an overlapping area. The new dimension for the smaller subsets is $d'/k$. The final dimension of each partition is $d'/k + o$, where $o$ is the overlapping size. \label{itemone}. The pseudocode for this process is shown in Algorithm~\ref{alg:normal}.
			
			\item \textbf{Subspacing of the shuffled genes into equal parts sequentially :} 
			\label{method:shuffle}
			\begin{algorithm}
				\caption{Sequential Shuffled Subspace}
				\label{alg:shuffle}
				\begin{algorithmic}[1]
					\Procedure{RandomizedSequentialDivision}{FeatureSet}
					\State Shuffle(FeatureSet)
					\State Define NumberOfPartitions, PartitionSize
					\State Initialize StartIndex to 0
					\While{StartIndex $<$ Length(FeatureSet)}
					\State EndIndex $\gets$ StartIndex + PartitionSize
					\State ShuffledDividedFeatureSets.add(FeatureSet[StartIndex:EndIndex])
					\State StartIndex $\gets$ StartIndex + (PartitionSize - OverlapSize)
					\EndWhile
					\EndProcedure
				\end{algorithmic}
			\end{algorithm}
			
			The partitioning approach-2 is nearly identical to the previous one (\ref{itemone}), except that, in this case, the features are randomly shuffled before creating the partitions. Overlapping is performed here with the same degree. The pseudocode for this process is shown in Algorithm~\ref{alg:shuffle}.
			
			\item \textbf{Subspacing based on random gene selection :} 
                \label{method:random}

			\begin{algorithm}
				\caption{Random Gene Selection}
				\label{alg:random}
				\begin{algorithmic}[1]
					\Procedure{DivideFeatures}{FeatureSet, k}
					\State Initialize $k$ empty divisions: $D_1, D_2, \ldots, D_k$
					\While{there are unselected genes in FeatureSet}
					\State Randomly select a gene $g$ from FeatureSet
					\State Randomly select a division $D_i$ from $D_1, D_2, \ldots, D_k$
					\If{$g \notin D_i$}
					\State Add $g$ to $D_i$
					\EndIf
					\EndWhile
					\State \textbf{return} $D_1, D_2, \ldots, D_k$
					\EndProcedure
				\end{algorithmic}
			\end{algorithm}
			
			In subspace generation approach-3, we create several buckets (or parts). Genes/features are then placed into these buckets through a random selection process. Specifically, genes are picked one by one at random and assigned to one of the randomly chosen buckets. This continues until every gene has been allocated to at least one bucket. A key rule is that a single bucket cannot contain the same gene more than once. However, different buckets can contain the same gene, allowing for overlap and ensuring that the distribution is similar to previous sections. This approach ensures a randomized yet controlled allocation of genes across multiple divisions. As shown in Algorithm~\ref{alg:random}, the procedure performs random gene selection.
			
			\item \textbf{Subspacing based on clustering of the genes }:
                \label{method:clsuter}
			
			\begin{algorithm}
				\caption{Cluster-Based Subspace}
				\label{alg:cluster}
				\begin{algorithmic}[1]
					\Procedure{ClusterBasedDivision}{FeatureSet}
					\State Define NumberOfClusters
					\State Clusters $\gets$ Cluster(FeatureSet, NumberOfClusters)
					\For{each ClusterIndex from 0 to NumberOfClusters-1}
					\State ClusteredDividedFeatureSets.add(Clusters[ClusterIndex])
					\EndFor
					\EndProcedure
				\end{algorithmic}
			\end{algorithm}
			
			 In approach 4, our algorithm employs the Leiden clustering algorithm to categorize genes into distinct groups. This method is referenced in the literature as Leiden\cite{Leiden}. We used the NETWORKX and IGRAPH packages in Python to construct the graph and the Leiden algorithm for clustering the features. Following the clustering process, the algorithm creates separate sections, each corresponding to one of the identified clusters. If the algorithm identifies 
            $k$ distinct clusters, then 
            $k$ separate divisions will be established. Within each division, genes that share similar characteristics are grouped together based on the clustering results. This organization ensures that each division is a collection of categorically similar genes, reflecting their shared cluster origin. The procedure that performs the clustering is shown in Algorithm~\ref{alg:cluster}
            		\end{enumerate}	
	
		\subsection{Data Extraction from different subspaces}
			
	   Principle component analysis(PCA) is a linear dimensionality reduction technique used for exploratory data analysis, visualization, and data preprocessing. It transforms the original data onto a new coordinate system, emphasizing directions (principal components) that capture the largest variation. The principal components are orthogonal unit vectors that form a basis where individual dimensions are linearly uncorrelated. It consists of mainly 5 stages, which are given below:
			\begin{enumerate}
				\item Standardizes the Range of Continuous Initial Variables, ensuring that all variables have the same scale (mean = 0, standard deviation = 1).
				\item Computes the Covariance Matrix to identify correlations between variables by calculating the covariance matrix.
				\item Computes Eigenvectors and Eigenvalues.
				\item Creates a Feature Vector to decide which principal components to keep based on their importance (determined by eigenvalues).	
				\item Finally, PCA transforms the data using the selected principal components.
			\end{enumerate}
			
	{\methodname} applies PCA on each subdivided section generated from the subspace generation process, leading to dimensionality reduction. PCA is performed to retain 95\% of the variance. As a result, the dimensionality of each section is reduced. As illustrated in Figure \ref{pic:pipeline}, the new dimensions for the gene axis obtained are m1, m2, ... and mk for k divisions.
			
		\subsection{Merging of the extracted datasets} 
			
		The {\methodname} algorithm then merges the k-divided partitions into a single datasets by concatenating them along the gene axis. As shown in Figure \ref{pic:pipeline}, the newly constructed dataset will have a dimension of  $m1 + m2 + ... + mk$, which is the sum of the dimensions of each section.
			
		\subsection{Clustering Algorithm Applied to the Merged Data}
	Any clustering algorithm can be used to determine clusters for our analysis. Our algorithm uses K-means clustering on the final merged dataset to determine the cluster information for each cell \cite{Kmeans_algo}.  We achieved remarkable results by applying the algorithm to the subspaced data. The clustering accuracy significantly improves when the input data is partitioned. These results are presented in Section~\ref{section:result}.
		

		

		\begin{table}[!ht]
			\centering
			\begin{adjustbox}{width=1\textwidth}
				\begin{tabular}{|l|l|l|l|l|}
					\hline
					Datasets Name & No. of Cells & No. of Genes & No. of Clusters & References \\ \hline
					Yan & 90 & 20214 & 7 & \cite{Yan} \\ \hline 
                    Yan2 & 90  & 20214  & 6 &  \cite{Intestine_data} \\ 
                    \hline
					Pollen & 301 & 23730  & 11 & \cite{Pollen} \\ \hline
                    Goolam & 124 & 41480   & 5 & \cite{Goolam} \\ \hline
                    Deng-rpkms & 268 & 22431   & 10  & \cite{Deng} \\ \hline
                    Fan & 69 & 26357   & 6 & \cite{Fan} \\ \hline
                    Kolod & 704 & 38653  & 3 & \cite{Kolod} \\ \hline
                    
				\end{tabular}
			\end{adjustbox}
			\caption{Summary of the scRNA-seq data used}
            \label{datasets}
		\end{table}

\section{Experimental Analysis}
\label{section:result}
All code used in the experimental analysis was developend using Python 3.8. We used Scikit-learn, Leidanlg, Tensorflow packages for our analysis. Codes are available here: \url{https://github.com/infiniteloop0048/FeatPCA}.
        
	\subsection{Datasets}
        The datasets used in our analyses are summarized in Table \ref{datasets}. \textit{Yan, Pollen, Goolam, Deng-rpkms, Fan and Kolod} datasets were downloaded from 
        \url{https://hemberg-lab.github.io/scRNA.seq.datasets/}{here}; and the Intestine dataset was downloaded from \url{http://cb.csail.mit.edu/cb/scanorama/data.tar.gz}.

\subsection{Performance Measurement}
		{\methodname} uses ARI value to find accuracy between clusters.
			
			\subsubsection{Adjusted Rand Index}
			The Adjusted Rand Index (ARI) is a measure used to evaluate the similarity between two data clusterings. It’s an adjustment of the Rand Index (RI) that corrects for the chance grouping of elements. It works the following way:
			
			\begin{itemize}
				\item \textbf{Rand Index (RI):} It’s a measure of the similarity between two clusterings by considering all pairs of samples and counting pairs that are assigned in the same or different clusters in the predicted and true clusterings. The RI score ranges from 0 (no pair classified in the same way under both clusterings) to 1 (all pairs are classified identically).
                Mathematically, the ARI value is calculated as follows - 
                \begin{equation}
                    R = \frac{{a + b}}{{nC2}}
                \end{equation}
                Where, $a$ represents the number of times a pair of elements belongs to the same cluster across both clustering methods. $b$ represents the number of times a pair of elements belongs to different clusters across both clustering methods. $nC2$ is the total number of unordered pairs in the dataset (where (n) is the number of elements).
				\\\\
				
				\item \textbf{Adjusted Rand Index (ARI):} Since the RI can be affected by chance, the ARI adjusts the RI by considering the expected index of agreement by chance. This makes the ARI more robust as it takes into account the possibility of random agreement. The ARI has a range of -1 to 1, where a value close to 1 indicates high similarity between clusterings, and a value close to 0 or negative indicates random or dissimilar clusterings.
				
			\end{itemize}
			
			Mathematically, the ARI value is calculated as follows - 
			\begin{equation}
				ARI = \frac{RI - Expected(RI)}{Max(RI) - Expected(RI)}
			\end{equation}

    \begin{table}[!ht]
        \centering
        \begin{adjustbox}{width=1\textwidth}
            \begin{tabular}{|p{3cm}|p{3cm}|p{3cm}|p{3cm}|}
                \hline
                Datasets Name & Sequencial Subspace & Shuffled sequential subspace & Subspace by random gene selection  \\ \hline
                Yan & 13/19 & 11/19 & 10/19  \\ \hline 
                Pollen & 15/19 & 3/19  & 3/19  \\ \hline
                Goolam & 10/19 & 6/19   & 13/19 \\ \hline
                Deng-rpkms & 16/19 & 12/19   & 12/19  \\ \hline
                Fan & 11/19 & 6/19   & 6/19  \\ \hline
                Kolod & 9/19 & 17/19  & 17/19  \\ \hline
                Intestine & 11/19  & 16/19  & 16/19  \\ \hline
            \end{tabular}
        \end{adjustbox}
        \caption{Win cases for the datasets. In the case of sequential subspace clustering for the “yan” dataset, the value “13/19” indicates that there are 13 instances where the ARI value improves with subspacing compared to using the undivided data.}
        \label{wincases}
    \end{table}

            \subsection{Results for the subspacing variations}
            \subsection*
			{Subspacing of the genes into equal parts sequencially}

	\begin{figure}
		\centering
		\begin{subfigure}{0.24\textwidth}
			\includegraphics[width=\linewidth]{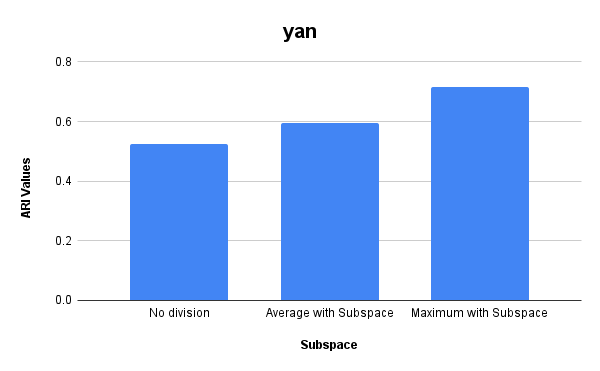}

		\end{subfigure}
		\hfill
		\begin{subfigure}{0.24\textwidth}
			\includegraphics[width=\linewidth]{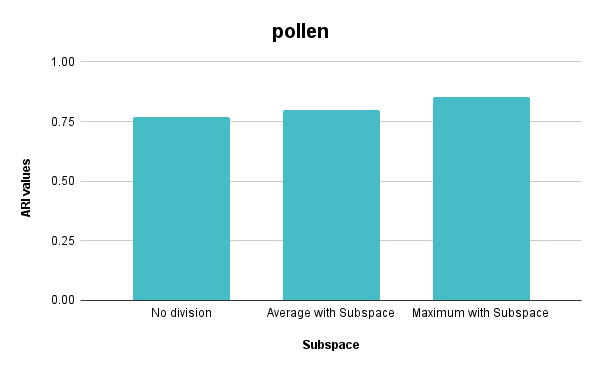}

		\end{subfigure}
		\hfill
		\begin{subfigure}{0.24\textwidth}
			\includegraphics[width=\linewidth]{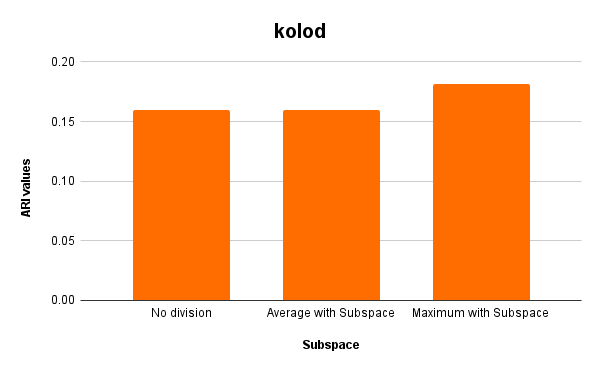}

		\end{subfigure}
		\hfill
		\begin{subfigure}{0.24\textwidth}
			\includegraphics[width=\linewidth]{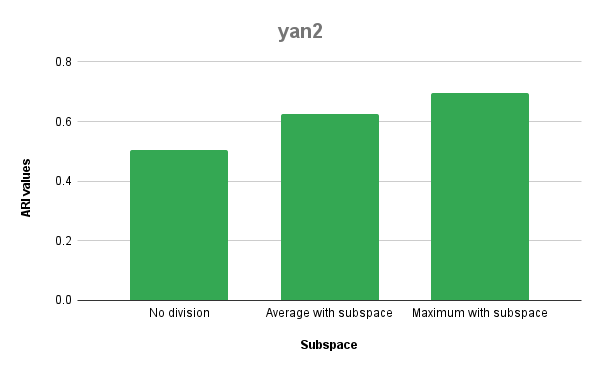}

		\end{subfigure}
		\hfill
		\begin{subfigure}{0.24\textwidth}
			\includegraphics[width=\linewidth]{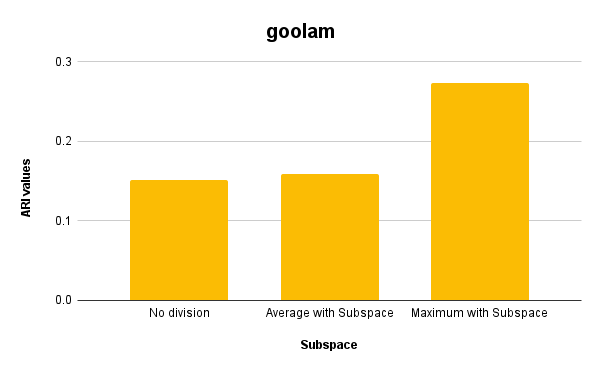}

		\end{subfigure}
		\hfill
		\begin{subfigure}{0.24\textwidth}
			\includegraphics[width=\linewidth]{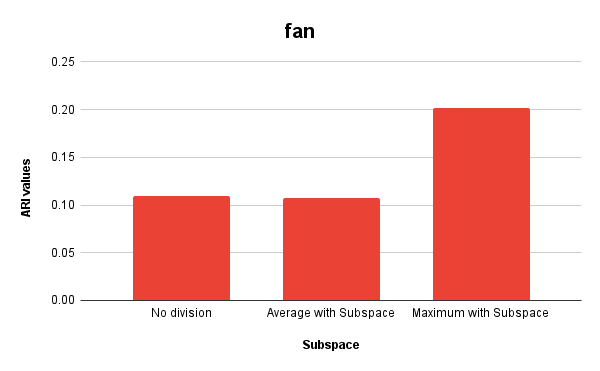}

		\end{subfigure}
		\hfill
		\begin{subfigure}{0.24\textwidth}
			\includegraphics[width=\linewidth]{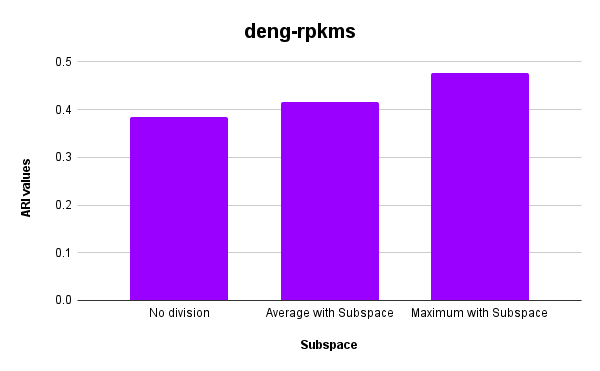}

		\end{subfigure}
		\caption{ARI value for sequencial subspacing}
		\label{pic:normal}
	\end{figure}

        Figure \ref{pic:normal} presents a series of seven bar plots, each corresponding to a distinct dataset. In every bar plot, 
	
	\begin{itemize}
		\item The initial bar represents the Adjusted Rand Index (ARI) value calculated for the undivided dataset.
		\item The second bar shows the average ARI across subspaces obtained using {\methodname}.
		\item The third bar indicates the maximum ARI among the subspaces.
	\end{itemize}
	
	The figure represents the Adjusted Rand Index (ARI) values for seven datasets when genes subspaceed sequencially. Across all datasets, the maximum ARI from subspaces consistently exceeds the ARI from the undivided dataset. Also for most of the datasets, the average Adjusted Rand Index (ARI) obtained from subspaces is superior. Notably, the ‘yan’ dataset achieves an ARI of 0.523 with the full dataset, but when divided into 14 subspaces, the maximum ARI reaches 0.716—a substantial improvement. Table \ref{wincases} shows the number of cases when subspacing produce better ARI values for the datasets.

			\subsection*
	{Subspacing of the shuffled genes into equal parts sequencially}
	
	\begin{figure}
		\centering
		\begin{subfigure}{0.24\textwidth}
			\includegraphics[width=\linewidth]{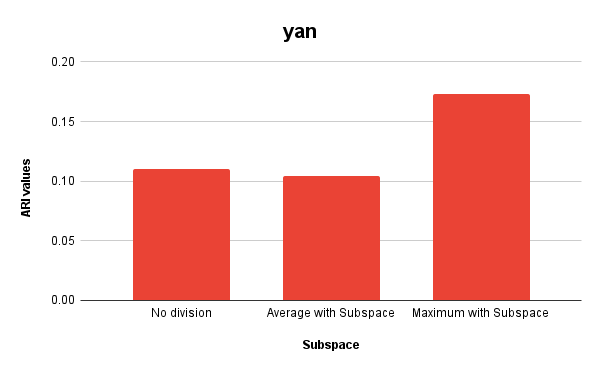}
			
		\end{subfigure}
		\hfill
		\begin{subfigure}{0.24\textwidth}
			\includegraphics[width=\linewidth]{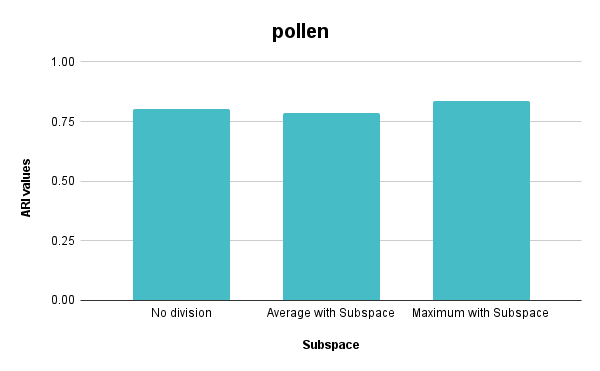}
			
		\end{subfigure}
		\hfill
		\begin{subfigure}{0.24\textwidth}
			\includegraphics[width=\linewidth]{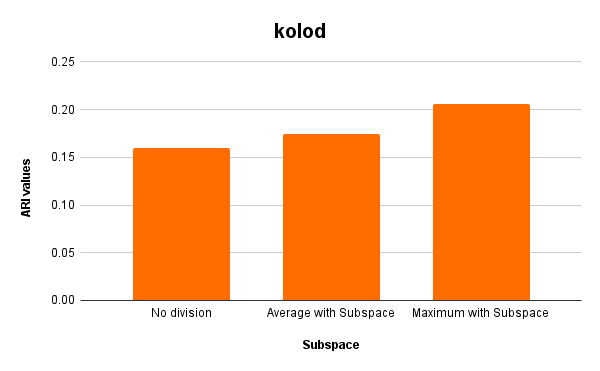}
			
		\end{subfigure}
		\hfill
		\begin{subfigure}{0.24\textwidth}
			\includegraphics[width=\linewidth]{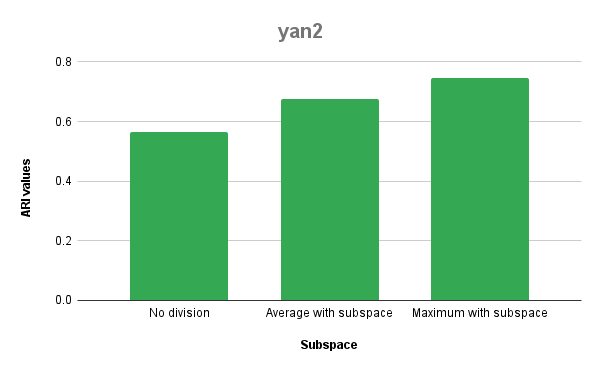}
			
		\end{subfigure}
		\hfill
		\begin{subfigure}{0.24\textwidth}
			\includegraphics[width=\linewidth]{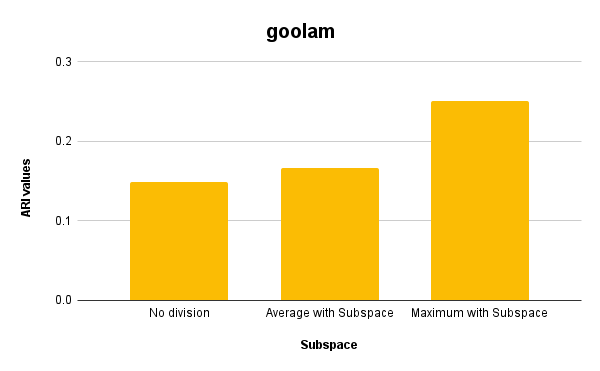}
			
		\end{subfigure}
		\hfill
		\begin{subfigure}{0.24\textwidth}
			\includegraphics[width=\linewidth]{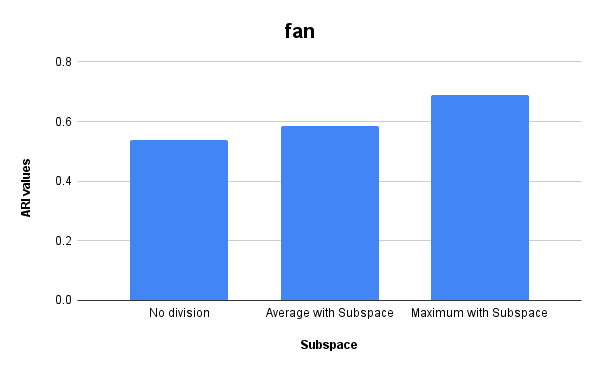}
			
		\end{subfigure}
		\hfill
		\begin{subfigure}{0.24\textwidth}
			\includegraphics[width=\linewidth]{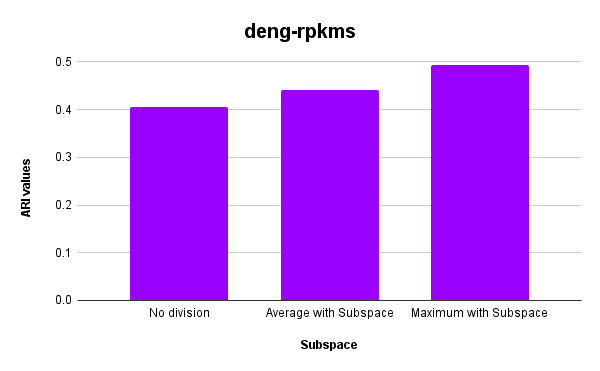}
			
		\end{subfigure}
		\caption{ARI value for shuffled sequencial subspacing based subspacing}
		\label{pic:shuffle}
	\end{figure}
		
			In Figure \ref{pic:shuffle}, the seven bar plots represent the Adjusted Rand Index (ARI) values for seven datasets when genes are shuffled and subspaced. Across all datasets, the maximum ARI from subspaces consistently surpasses the ARI from the undivided dataset. With the exception of the ‘yan’ dataset, the average ARI obtained from subspaces is superior. Notably, the ‘pollen’ dataset achieves an ARI of 0.802 with the full dataset, but when divided into three subspaces, the maximum ARI reaches 0.835—a substantial improvement. Table \ref{wincases} displays the count of cases where subspacing results in improved ARI values for the datasets.

			\subsection*
			{Subspacing based on random gene selection}
			
			\begin{figure}
		\centering
		\begin{subfigure}{0.24\textwidth}
			\includegraphics[width=\linewidth]{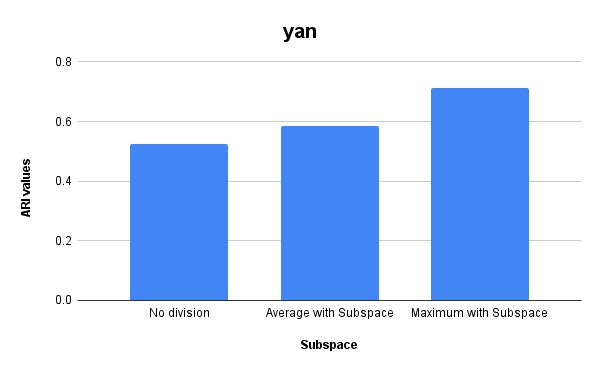}
			
		\end{subfigure}
		\hfill
		\begin{subfigure}{0.24\textwidth}
			\includegraphics[width=\linewidth]{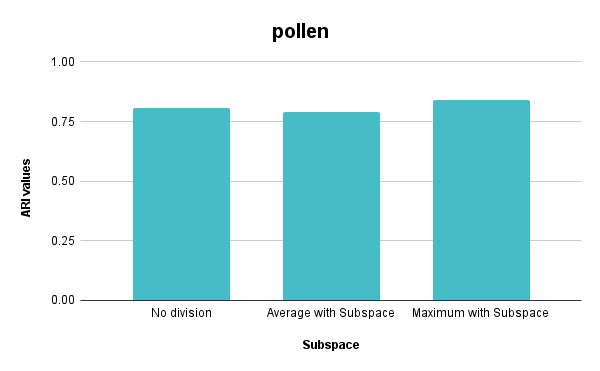}
			
		\end{subfigure}
		\hfill
		\begin{subfigure}{0.24\textwidth}
			\includegraphics[width=\linewidth]{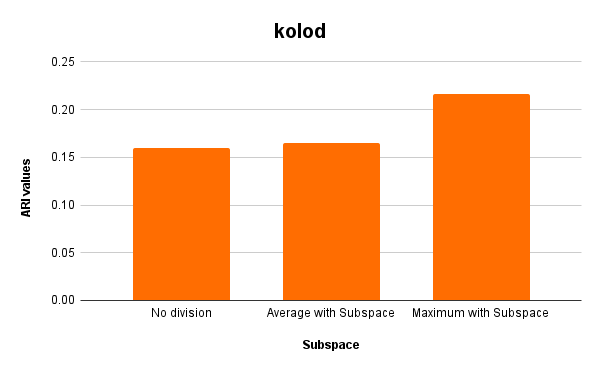}
			
		\end{subfigure}
		\hfill
		\begin{subfigure}{0.24\textwidth}
			\includegraphics[width=\linewidth]{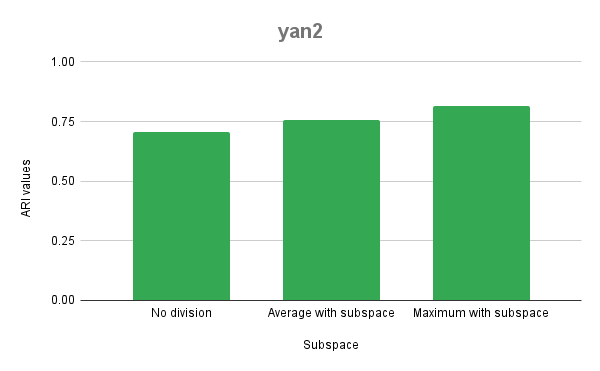}
			
		\end{subfigure}
		\hfill
		\begin{subfigure}{0.24\textwidth}
			\includegraphics[width=\linewidth]{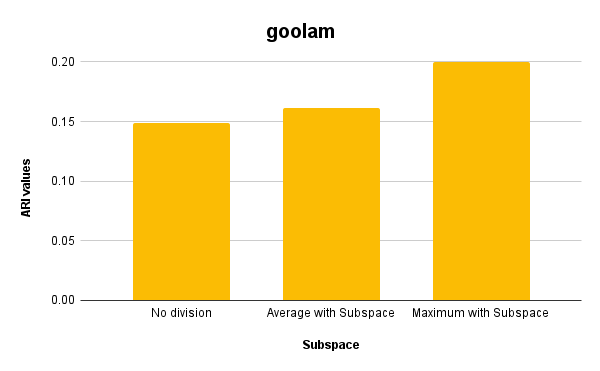}
			
		\end{subfigure}
		\hfill
		\begin{subfigure}{0.24\textwidth}
			\includegraphics[width=\linewidth]{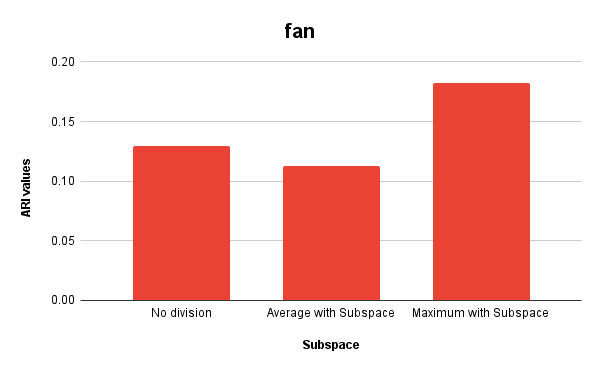}
			
		\end{subfigure}
		\hfill
		\begin{subfigure}{0.24\textwidth}
			\includegraphics[width=\linewidth]{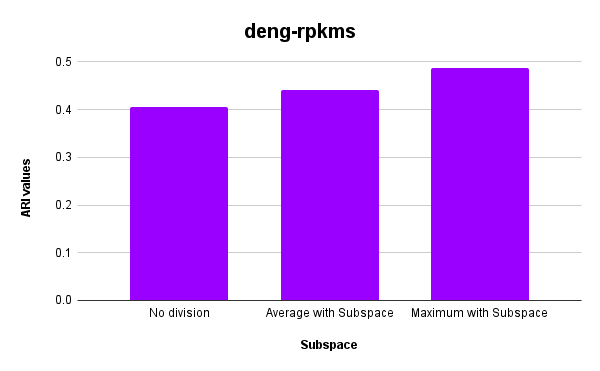}
			
		\end{subfigure}
		\caption{ARI value for randomly selected genes}
		\label{pic:bucket}
	\end{figure}
			
			Figure \ref{pic:bucket} depicts the adjusted Rand index (ARI) values obtained by randomly selecting genes and dividing them into partitions. Notably, specific divisions consistently yield higher ARI values compared to using the entire dataset. For the “deng-rpkms” dataset, the maximum ARI achieved from subspaces is 0.479, surpassing the ARI of 0.406 obtained when the dataset remains undivided. Additionally, the average ARI from the subspaces generally outperforms that of most other datasets. Table \ref{wincases} provides the count of cases where subspacing results in improved ARI values for the datasets.

			\subsection*
			{Subsapcing based on clustering of the genes}
			
				\begin{figure}
		\centering
		\begin{subfigure}{0.24\textwidth}
			\includegraphics[width=\linewidth]{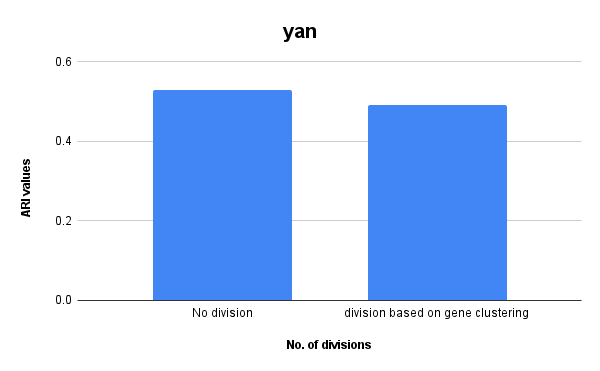}
			
		\end{subfigure}
		\hfill
		\begin{subfigure}{0.24\textwidth}
			\includegraphics[width=\linewidth]{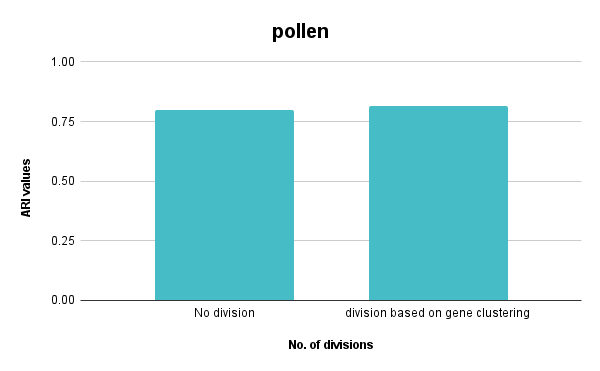}
			
		\end{subfigure}
		\hfill
		\begin{subfigure}{0.24\textwidth}
			\includegraphics[width=\linewidth]{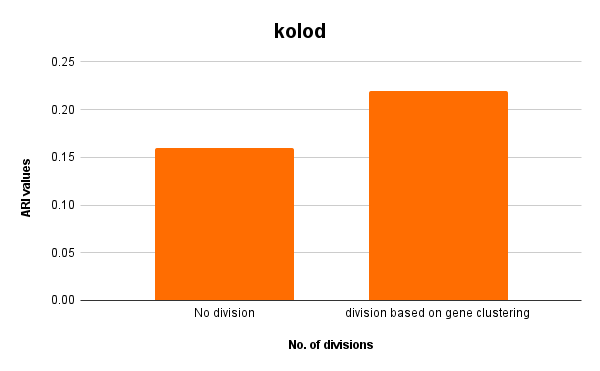}
			
		\end{subfigure}
		\hfill
		\begin{subfigure}{0.24\textwidth}
			\includegraphics[width=\linewidth]{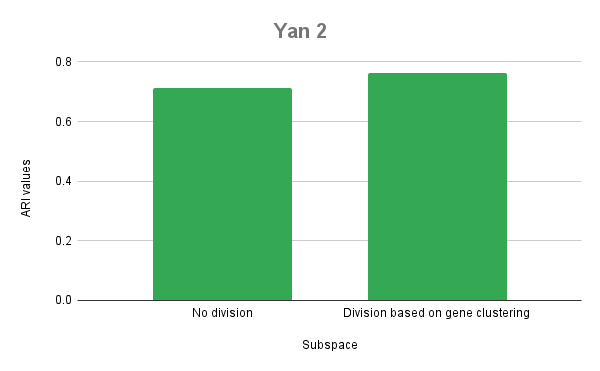}
			
		\end{subfigure}
		\hfill
		\begin{subfigure}{0.24\textwidth}
			\includegraphics[width=\linewidth]{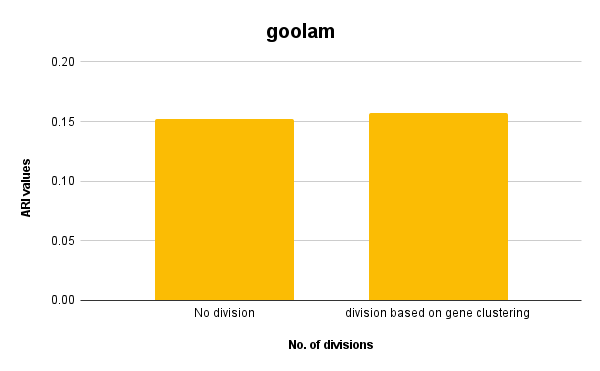}
			
		\end{subfigure}
		\hfill
		\begin{subfigure}{0.24\textwidth}
			\includegraphics[width=\linewidth]{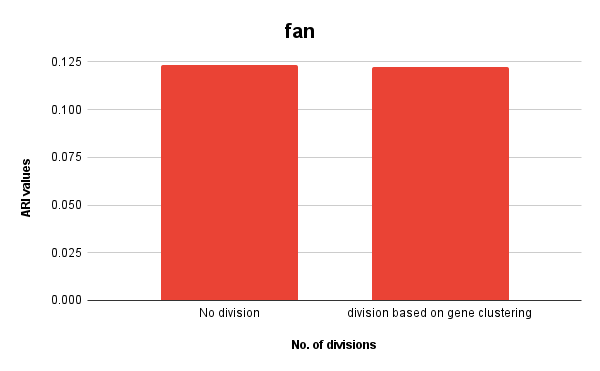}
			
		\end{subfigure}
		\hfill
		\begin{subfigure}{0.24\textwidth}
			\includegraphics[width=\linewidth]{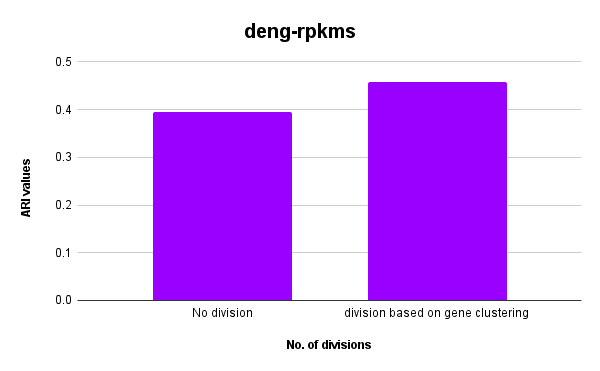}
			
		\end{subfigure}
		\caption{ARI value based on performing gene clustering}
		\label{pic:cluster}
	\end{figure}

			Figure \ref{pic:cluster} illustrates the ARI values when the genes are divided according to gene clustering. The figure comprises seven barplots, each corresponding to one of the seven datasets. In each barplot, the first bar represents the Adjusted Rand Index (ARI) value for the entire dataset, while the second bar indicates the ARI value for the subspace dataset based on feature clustering. The Figure shows that for 5 out of 7 datasets, certain divisions yield a higher ARI value compared to using the full dataset. For instance, for the "intestine" dataset, the ARI value is 0.440 when using the full dataset (no division). However, when the genes are divided, the ARI value increases to 0.574, significantly higher than that of the undivided data. Similarly, improved results are observed for other datasets when the data is divided, as illustrated in the Figure.
	\subsection{Comparison with other state-of-the-art method}
	
	In our study, we conducted a comparative analysis by benchmarking our results against three state-of-the-art methods: SC3, Seurat, and FEATS. In our study, we conducted a comparative analysis by benchmarking our results against three state-of-the-art methods: SC3, Seurat, and FEATS. In our comparison, we initially imputed missing gene expression values using an autoencoder. Specifically, we replaced the missing entries with values obtained from the autoencoder model. In the \textit{yan} dataset, the highest Adjusted Rand Index (ARI) value (0.8938) was achieved using the \hyperref[method:random]{random gene selection subspacing} method. For the \textit{pollen, goolam, deng, and kolod} datasets, the maximum ARI values were obtained using the \hyperref[method:shuffle]{shuffled subspacing} method. In the case of the \textit{fan} datasets, the maximum ARI value was obtained by applying the \hyperref[method:normal]{sequencial subspacing} method. Remarkably, our method exhibited superior performance across nearly all cases. ARI values of all the methods are given in Table  \ref{comparison}.

	\begin{table}[!htb]
		\centering
		\begin{adjustbox}{width=1\textwidth}
			\begin{tabular}{|p{3cm}|p{3cm}|p{3cm}|p{3cm}|p{3cm}|}
				\hline
				Datasets Name & {\methodname} & SC3{\cite{sc3}} & Seurat{\cite{seurat}} & Feats{\cite{feats}}  \\ \hline
				Yan & \textbf{0.8938} & 0.6584 & 0.6297  & 0.8029\\ \hline 
                Yan 2 & \textbf{0.8144} & 0.6549 & 0.6408  & 0.7421\\
                \hline
				Pollen & 0.9142 & 0.9581  & 0.8420 & \textbf{0.9754}\\ \hline
				Goolam & \textbf{0.9561} & 0.5441  & 0.3102 & 0.8518\\ \hline
				Deng-rpkms & 0.4457 & \textbf{0.6659}   & 0.5805 & 0.4562\\ \hline
				Fan & \textbf{0.3692} & 0.3442   & 0.2503  & 0.3677\\ \hline
				Kolod & 0.8111 & 0.1474  & 0.5885  & \textbf{0.9920}\\ \hline
               
			\end{tabular}
		\end{adjustbox}
		\caption{Comparing the Adjusted Rand Index (ARI) values of existing clustering methods with {\methodname}. For the intestine dataset, no results were obtained from Feats.}
		\label{comparison}
	\end{table}
 \vspace{-1cm}
\section{Conclusion}	
		 In our study, we explored the impact of data subspacing along feature sets on clustering accuracy. Notably, dividing the data into subspaces based on relevant features significantly improved clustering results. Specifically, applying Principal Component Analysis (PCA) to these subspaces outperformed applying PCA to the entire dataset when identifying distinct cell types. Among the four methods we investigated, sequential subspacing emerged as the most successful approach, yielding the highest number of “win cases" instances where subspacing improved the Adjusted Rand Index (ARI). However, for datasets with relatively fewer clusters, subspacing via random gene selection also demonstrated promising results. We systematically divided the data into 2 to 20 partitions, consistently observing better outcomes across most divisions. Our future work will delve deeper into understanding the relationship between the number of subspaces and clustering performance. It would be fascinating to predict the optimal subspaces count that maximizes ARI and explore how changes in the number of subspaces impact clustering accuracy. We will also extend our focus to multimodal data and apply the same principles. Our method is independent of the specific clustering and dimension reduction algorithms. However, in this analysis, we utilize PCA for dimensionality reduction and K-means for clustering. In summary, our findings emphasize the importance of thoughtful feature partitioning while clustering.
		\appendix

		\bibliography{references}

\begin{thebibliography}{10}

\bibitem{seq1}
Malte~D Luecken and Fabian~J Theis.
\newblock Current best practices in single-cell rna-seq analysis: a tutorial.
\newblock {\em Molecular systems biology}, 15(6):e8746, 2019.

\bibitem{seq2}
Geng Chen, Baitang Ning, and Tieliu Shi.
\newblock Single-cell rna-seq technologies and related computational data analysis.
\newblock {\em Frontiers in genetics}, 10:317, 2019.

\bibitem{applications1}
Yalan Lei, Rong Tang, Jin Xu, Wei Wang, Bo~Zhang, Jiang Liu, Xianjun Yu, and Si~Shi.
\newblock Applications of single-cell sequencing in cancer research: progress and perspectives.
\newblock {\em Journal of hematology \& oncology}, 14(1):91, 2021.

\bibitem{applications2}
Alev Baysoy, Zhiliang Bai, Rahul Satija, and Rong Fan.
\newblock The technological landscape and applications of single-cell multi-omics.
\newblock {\em Nature Reviews Molecular Cell Biology}, 24(10):695--713, 2023.

\bibitem{applications3}
Anton Bankevich, Sergey Nurk, Dmitry Antipov, Alexey~A Gurevich, Mikhail Dvorkin, Alexander~S Kulikov, Valery~M Lesin, Sergey~I Nikolenko, Son Pham, Andrey~D Prjibelski, et~al.
\newblock Spades: a new genome assembly algorithm and its applications to single-cell sequencing.
\newblock {\em Journal of computational biology}, 19(5):455--477, 2012.

\bibitem{missingvalues}
Stephanie~C Hicks, F~William Townes, Mingxiang Teng, and Rafael~A Irizarry.
\newblock Missing data and technical variability in single-cell rna-sequencing experiments.
\newblock {\em Biostatistics}, 19(4):562--578, 2018.

\bibitem{imputing}
Lihua Zhang and Shihua Zhang.
\newblock Comparison of computational methods for imputing single-cell rna-sequencing data.
\newblock {\em IEEE/ACM transactions on computational biology and bioinformatics}, 17(2):376--389, 2018.

\bibitem{autoimpute}
Divyanshu Talwar, Aanchal Mongia, Debarka Sengupta, and Angshul Majumdar.
\newblock Autoimpute: Autoencoder based imputation of single-cell rna-seq data.
\newblock {\em Scientific reports}, 8(1):16329, 2018.

\bibitem{dimred1}
Koki Tsuyuzaki, Hiroyuki Sato, Kenta Sato, and Itoshi Nikaido.
\newblock Benchmarking principal component analysis for large-scale single-cell rna-sequencing.
\newblock {\em Genome biology}, 21(1):9, 2020.

\bibitem{dimred2}
Snehalika Lall, Debajyoti Sinha, Sanghamitra Bandyopadhyay, and Debarka Sengupta.
\newblock Structure-aware principal component analysis for single-cell rna-seq data.
\newblock {\em Journal of Computational Biology}, 25(12):1365--1373, 2018.

\bibitem{clustering1}
Raphael Petegrosso, Zhuliu Li, and Rui Kuang.
\newblock Machine learning and statistical methods for clustering single-cell rna-sequencing data.
\newblock {\em Briefings in bioinformatics}, 21(4):1209--1223, 2020.

\bibitem{clustering2}
Yanglan Gan, Ning Li, Guobing Zou, Yongchang Xin, and Jihong Guan.
\newblock Identification of cancer subtypes from single-cell rna-seq data using a consensus clustering method.
\newblock {\em BMC medical genomics}, 11:65--72, 2018.

\bibitem{kmeans}
Liang Chen, Weinan Wang, Yuyao Zhai, and Minghua Deng.
\newblock Deep soft k-means clustering with self-training for single-cell rna sequence data.
\newblock {\em NAR genomics and bioinformatics}, 2(2):lqaa039, 2020.

\bibitem{dimensionality}
Md~Rezaul Karim, Oya Beyan, Achille Zappa, Ivan~G Costa, Dietrich Rebholz-Schuhmann, Michael Cochez, and Stefan Decker.
\newblock Deep learning-based clustering approaches for bioinformatics.
\newblock {\em Briefings in bioinformatics}, 22(1):393--415, 2021.

\bibitem{dimensionality1}
Bahjat~F Qaqish, Jonathon~J O’Brien, Jonathan~C Hibbard, and Katie~J Clowers.
\newblock {Accelerating high-dimensional clustering with lossless data reduction}.
\newblock {\em Bioinformatics}, 33(18):2867--2872, 05 2017.

\bibitem{umap}
Leland McInnes, John Healy, and James Melville.
\newblock Umap: Uniform manifold approximation and projection for dimension reduction.
\newblock {\em arXiv preprint arXiv:1802.03426}, 2018.

\bibitem{tsne}
Laurens Van~der Maaten and Geoffrey Hinton.
\newblock Visualizing data using t-sne.
\newblock {\em Journal of machine learning research}, 9(11), 2008.

\bibitem{lda}
Ronald~A Fisher.
\newblock The use of multiple measurements in taxonomic problems.
\newblock {\em Annals of eugenics}, 7(2):179--188, 1936.

\bibitem{autoencoders}
Mayu Sakurada and Takehisa Yairi.
\newblock Anomaly detection using autoencoders with nonlinear dimensionality reduction.
\newblock In {\em Proceedings of the MLSDA 2014 2nd workshop on machine learning for sensory data analysis}, pages 4--11, 2014.

\bibitem{d1}
Yingyu Liang, Maria-Florina~F Balcan, Vandana Kanchanapally, and David Woodruff.
\newblock Improved distributed principal component analysis.
\newblock {\em Advances in neural information processing systems}, 27, 2014.

\bibitem{d2}
Yingyu Liang, Maria-Florina Balcan, and Vandana Kanchanapally.
\newblock Distributed pca and k-means clustering.
\newblock In {\em The Big Learning Workshop at NIPS}, volume 2013. Citeseer, 2013.

\bibitem{autoimpute1}
Md~Bahadur Badsha, Rui Li, Boxiang Liu, Yang~I Li, Min Xian, Nicholas~E Banovich, and Audrey Qiuyan~Fu.
\newblock Imputation of single-cell gene expression with an autoencoder neural network.
\newblock {\em Quantitative Biology}, 8(1):78--94, 2020.

\bibitem{pcaapplication}
Federico Marini and Harald Binder.
\newblock pcaexplorer: an r/bioconductor package for interacting with rna-seq principal components.
\newblock {\em BMC bioinformatics}, 20:1--8, 2019.

\bibitem{umapapplication}
Etienne Becht, Leland McInnes, John Healy, Charles-Antoine Dutertre, Immanuel~WH Kwok, Lai~Guan Ng, Florent Ginhoux, and Evan~W Newell.
\newblock Dimensionality reduction for visualizing single-cell data using umap.
\newblock {\em Nature biotechnology}, 37(1):38--44, 2019.

\bibitem{tsneapplication}
George~C Linderman, Manas Rachh, Jeremy~G Hoskins, Stefan Steinerberger, and Yuval Kluger.
\newblock Fast interpolation-based t-sne for improved visualization of single-cell rna-seq data.
\newblock {\em Nature methods}, 16(3):243--245, 2019.

\bibitem{autoencoderapplication}
Dongfang Wang and Jin Gu.
\newblock Vasc: dimension reduction and visualization of single-cell rna-seq data by deep variational autoencoder.
\newblock {\em Genomics, Proteomics and Bioinformatics}, 16(5):320--331, 2018.

\bibitem{datap1}
Ahmad~B Hassanat, Ahmad~S Tarawneh, Samer~Subhi Abed, Ghada~Awad Altarawneh, Malek Alrashidi, and Mansoor Alghamdi.
\newblock Rdpvr: Random data partitioning with voting rule for machine learning from class-imbalanced datasets.
\newblock {\em Electronics}, 11(2):228, 2022.

\bibitem{d3}
Hillol Kargupta, Weiyun Huang, Krishnamoorthy Sivakumar, and Erik Johnson.
\newblock Distributed clustering using collective principal component analysis.
\newblock {\em Knowledge and Information Systems}, 3:422--448, 2001.

\bibitem{scanpy}
F.~Wolf, Philipp Angerer, and Fabian Theis.
\newblock Scanpy: Large-scale single-cell gene expression data analysis.
\newblock {\em Genome Biology}, 19, 02 2018.

\bibitem{Leiden}
Vincent~A Traag, Ludo Waltman, and Nees~Jan Van~Eck.
\newblock From louvain to leiden: guaranteeing well-connected communities.
\newblock {\em Scientific reports}, 9(1):1--12, 2019.

\bibitem{Kmeans_algo}
T~Kanungo, D~M Mount, N~S Netanyahu, C~D Piatko, R~Silverman, and A~Y Wu.
\newblock An efficient k-means clustering algorithm: analysis and implementation.
\newblock {\em IEEE Trans. Pattern Anal. Mach. Intell.}, 24(7):881--892, July 2002.

\bibitem{Yan}
Liying Yan, Mingyu Yang, Hongshan Guo, Lu~Yang, Jun Wu, Rong Li, Ping Liu, Ying Lian, Xiaoying Zheng, Jie Yan, Jin Huang, Ming Li, Xinglong Wu, Lu~Wen, Kaiqin Lao, Ruiqiang Li, Jie Qiao, and Fuchou Tang.
\newblock Single-cell {RNA-Seq} profiling of human preimplantation embryos and embryonic stem cells.
\newblock {\em Nat. Struct. Mol. Biol.}, 20(9):1131--1139, September 2013.

\bibitem{Intestine_data}
Toshiro Sato, Robert~G Vries, Hugo~J Snippert, Marc van~de Wetering, Nick Barker, Daniel~E Stange, Johan~H van Es, Arie Abo, Pekka Kujala, Peter~J Peters, and Hans Clevers.
\newblock Single lgr5 stem cells build crypt-villus structures in vitro without a mesenchymal niche.
\newblock {\em Nature}, 459(7244):262--265, May 2009.

\bibitem{Pollen}
Alex~A Pollen, Tomasz~J Nowakowski, Joe Shuga, Xiaohui Wang, Anne~A Leyrat, Jan~H Lui, Nianzhen Li, Lukasz Szpankowski, Brian Fowler, Peilin Chen, Naveen Ramalingam, Gang Sun, Myo Thu, Michael Norris, Ronald Lebofsky, Dominique Toppani, Darnell~W Kemp, 2nd, Michael Wong, Barry Clerkson, Brittnee~N Jones, Shiquan Wu, Lawrence Knutsson, Beatriz Alvarado, Jing Wang, Lesley~S Weaver, Andrew~P May, Robert~C Jones, Marc~A Unger, Arnold~R Kriegstein, and Jay A~A West.
\newblock Low-coverage single-cell {mRNA} sequencing reveals cellular heterogeneity and activated signaling pathways in developing cerebral cortex.
\newblock {\em Nat. Biotechnol.}, 32(10):1053--1058, October 2014.

\bibitem{Goolam}
Mubeen Goolam, Antonio Scialdone, Sarah J~L Graham, Iain~C Macaulay, Agnieszka Jedrusik, Anna Hupalowska, Thierry Voet, John~C Marioni, and Magdalena Zernicka-Goetz.
\newblock Heterogeneity in oct4 and sox2 targets biases cell fate in 4-cell mouse embryos.
\newblock {\em Cell}, 165(1):61--74, March 2016.

\bibitem{Deng}
Qiaolin Deng, Daniel Ramsköld, Björn Reinius, and Rickard Sandberg.
\newblock Single-cell rna-seq reveals dynamic, random monoallelic gene expression in mammalian cells.
\newblock {\em Science}, 343(6167):193--196, 2014.

\bibitem{Fan}
Xiaoying Fan, Xiannian Zhang, Xinglong Wu, Hongshan Guo, Yuqiong Hu, Fuchou Tang, and Yanyi Huang.
\newblock Single-cell {RNA-seq} transcriptome analysis of linear and circular {RNAs} in mouse preimplantation embryos.
\newblock {\em Genome Biol.}, 16(1):148, July 2015.

\bibitem{Kolod}
Aleksandra~A Kolodziejczyk, Jong~Kyoung Kim, Jason C~H Tsang, Tomislav Ilicic, Johan Henriksson, Kedar~N Natarajan, Alex~C Tuck, Xuefei Gao, Marc B{\"u}hler, Pentao Liu, John~C Marioni, and Sarah~A Teichmann.
\newblock Single cell {RNA-sequencing} of pluripotent states unlocks modular transcriptional variation.
\newblock {\em Cell Stem Cell}, 17(4):471--485, October 2015.

\bibitem{sc3}
Vladimir~Yu Kiselev, Kristina Kirschner, Michael~T Schaub, Tallulah Andrews, Andrew Yiu, Tamir Chandra, Kedar~N Natarajan, Wolf Reik, Mauricio Barahona, Anthony~R Green, et~al.
\newblock Sc3: consensus clustering of single-cell rna-seq data.
\newblock {\em Nature methods}, 14(5):483--486, 2017.

\bibitem{seurat}
Tim Stuart, Andrew Butler, Paul Hoffman, Christoph Hafemeister, Efthymia Papalexi, William~M Mauck, Yuhan Hao, Marlon Stoeckius, Peter Smibert, and Rahul Satija.
\newblock Comprehensive integration of single-cell data.
\newblock {\em cell}, 177(7):1888--1902, 2019.

\bibitem{feats}
Edwin Vans, Ashwini Patil, and Alok Sharma.
\newblock Feats: feature selection-based clustering of single-cell rna-seq data.
\newblock {\em Briefings in Bioinformatics}, 22(4):bbaa306, 2021.

\end{thebibliography}
		\bibliographystyle{unsrt}

\end{document}